\documentclass[prl,twocolumn,showpacs,superscriptaddress,showkeys]{revtex4}
\usepackage{graphicx}
\usepackage{dcolumn}
\usepackage{bm}
\begin{document}

\title{Generation of High Quality Laser Accelerated Ion Beams}

\author{T.~Zh.~Esirkepov}
 \altaffiliation[Also at ]{Moscow Institute of Physics and Technology, Dolgoprudny, Moscow region 141700, Russia}
 \email{timur@ile.osaka-u.ac.jp}
\affiliation{Institute of Laser Engineering, Osaka University, Osaka 565-0871, Japan}

\author{S.~V.~Bulanov}
 \altaffiliation[Also at ]{General Physics Institute of RAS, Moscow 119991, Russia}
\affiliation{Advanced Photon Research Center, JAERI, Kyoto-fu 619-0215, Japan}

\author{K.~Nishihara}
\affiliation{Institute of Laser Engineering, Osaka University, Osaka 565-0871, Japan}

\author{T.~Tajima}
\affiliation{Advanced Photon Research Center, JAERI, Kyoto-fu 619-0215, Japan}

\author{F.~Pegoraro}
\affiliation{University of Pisa and INFM, Pisa 56100, Italy}

\author{V.~S.~Khoroshkov}
\affiliation{Institute of Theoretical and Experimental Physics, Moscow 117259, Russia}

\author{K.~Mima}
\affiliation{Institute of Laser Engineering, Osaka University, Osaka 565-0871, Japan}

\author{H.~Daido}
\author{Y.~Kato}
\affiliation{Advanced Photon Research Center, JAERI, Kyoto-fu 619-0215, Japan}

\author{Y.~Kitagawa}
\author{K.~Nagai}
\author{S.~Sakabe}
\affiliation{Institute of Laser Engineering, Osaka University, Osaka 565-0871, Japan}

\date{June 16, 2002} 

\begin{abstract}
In order to achieve
a high quality, i. e. monoergetic, intense ion beam,
we propose the use of a double layer target.
The first layer, at the target front, consists of high-Z atoms,
while the second (rear) layer is a thin coating of low-Z atoms.
The high quality proton beams from the double layer target,
irradiated by an ultra-intense laser pulse,
are  demonstrated with three dimensional Particle-in-Cell simulations.
\end{abstract}

\pacs{52.50.Jm, 52.59.-f, 52.65.Rr}

\keywords{Ion acceleration, monoenergetic ion beams,
laser plasma interaction, Particle-in-Cell simulation}

\maketitle

\begin{figure}
\includegraphics{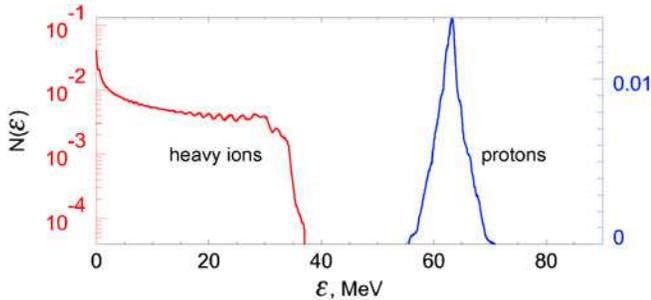}
\caption{\label{fig:E}
Electric field near the target is shown as the 3D vector field,
the length of each vector corresponds to the magnitude of electric field.
Vectors with $|eE_x/(m_e\omega c)|\le 1$ are violet.
The laser pulse is presented by isosurfaces of the transverse component
$E_z$ corresponding to dimensionless values $\pm 10$.
The time unit is the laser period $2\pi /\omega$.}
\end{figure}

\begin{figure}
\includegraphics{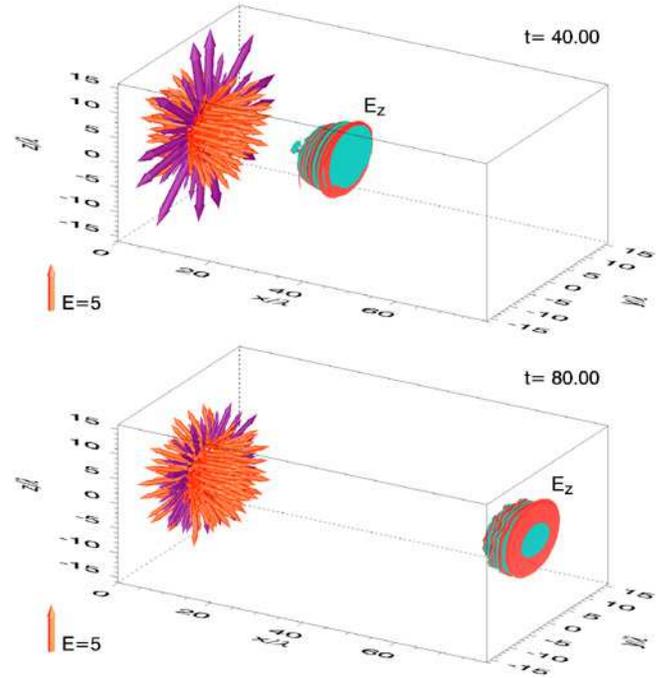}
\caption{\label{fig:Z}
Plasma species.
The envelopes of
heavy ions (thick red shell) and
light ions (thin blue shell) are shown.
Electron density is shown as a `green gas'
using ray tracing technique.}
\end{figure}

\begin{figure}
\includegraphics{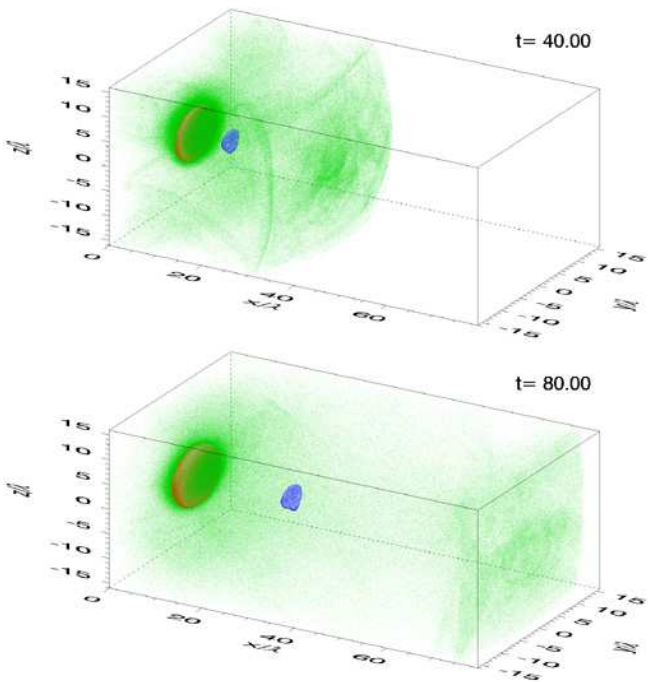}
\caption{\label{fig:spectra}
The proton (blue) and heavy ion (red) energy spectra at $t=80\cdot 2\pi /\omega$.}
\end{figure}

The  high efficiency of ion acceleration recently observed in the
interaction of petawatt laser pulses \cite{Mourou98}  with solid
targets \cite{ions1} has  lead to  important applications such as
the use of laser produced fast ion beam injection to conventional
accelerators (see Ref. \cite{injector-00}), the fast ignition of
thermonuclear targets, as discussed in Refs. \cite{Roth01}, and
 hadrontherapy in oncology
\cite{BKh}.
Laser accelerated protons have been used
for proton imaging of  small scale objects
in  laser produced plasmas
with high time resolution \cite{Borghesi}.
The process of ion
acceleration has been studied in detail with multi-dimensional
Particle-in-Cell (PIC) simulations \cite{Es99}.
In the recent
experimental results presented in Refs. \cite{ions1},
electron
energy  in a range of hundreds of MeV was observed, while
the proton energy was about tens of MeV with a  number of fast
protons from $10^{12}$ to $10^{13}$ per pulse and with the transformation of 
$12\%$ of the laser energy into fast
ion energy. 
In Refs. \cite{Es99} it was shown with PIC simulations
that, by optimizing the laser-target parameters, it becomes
possible to accelerate protons up to several
hundreds MeV with a number of fast ions approximately
$10^{13}$ particles per pulse.

The typical energy spectrum of laser accelerated particles
observed  both in the experiments and in the computer simulations
can be approximated by a quasi-thermal distribution with a cut-off
at a maximum energy ${\cal E}_{\max}$.
The effective temperature
$T$, that may be attributed to the fast ion beams,
is much less than the maximal energy.
On the other hand, almost all  above mentioned applications
require high quality proton beams, i.e.,  beams with sufficiently
small energy spread $\Delta {\cal E}/{\cal E}$.
For example, for the hadron therapy it is highly desirable
to have a proton beam with
$\Delta {\cal E}/{\cal E}\leq 2\%$ in order to provide the
conditions for a high irradiation dose being delivered to the
tumor saving neighboring healthy tissues \cite{KhM}.
In the concept of Fast Ignition with laser accelerated ions presented in
Refs. \cite{Roth01}, the proton beam was assumed to be
quasi-monoenergetic.
An analysis carried out in Ref. \cite{Atzeni}
has shown that the Fast Ignition with a
quasi-thermal beam of fast protons requires
several times larger energy
than that with a monoenergetic beam.
Similarly, in the case of the injector (see
Ref. \cite{injector-00}), a high-quality beam is needed in order
to inject the charged particles into  the optimal
accelerating phase efficiently.
Thus we see that the  generation  of high
quality beams is a key problem for many applications.
However
energy spectra of the laser accelerated ions at present are rather
far from required ones.

In this Letter we show with three dimensional (3D) PIC simulations that such a
required beam of laser accelerated ions can be obtained using a
double layer target (see also Refs. \cite{BKh}). Multi layer
targets have been used for a long time in order to increase the
efficiency of the laser energy conversion into plasma and fast
particle kinetic energy (see Refs. in \cite{Double}). In contrast to
the configurations discussed previously, we propose to use
a double layer target to produce fast proton beams with controlled
quality.

In the proposed scheme the target is made of two layers.
The first layer consists of high-Z atoms (atomic mass $m_i$),
while the second layer is
a very thin coating of low-Z atoms (atomic mass $m_a$).
Such the target can be a metal foil
coated with a thin hydrogeneous film.
An ultra-intense laser pulse is incident on the first layer,
and so we say that the first layer is at the target front,
while the second layer is at the rear side of the target.
We use the term `longitudinal' for the direction of
propagation of the laser pulse, and the term `transverse'
for perpendicular directions.

When an ultra-intense laser pulse irradiates the target,
heavy atoms are partly ionized and
electrons are expelled from the foil.
A quasi-static electric field is generated due to charge separation.
The first layer of heavy ions (the foil)
should be sufficiently thick so as to produce a large enough
quasi-static electric field, and, at the same time,
it should be sufficiently thin to be able to produce
strong electric field at the rear side.
Such an electric field has
opposite signs on the two sides of the target,
and vanishes inside the target and at some finite distance from it.
The number of low-Z ions in the second layer (the coating)
should be sufficiently small not to
produce any significant effect on the electric field.
The quasi-static electric field accelerates
both high-Z ions (with average charge $e Z_i$) and
low-Z ions (with average charge $e Z_a$).
If the ratio $m_i Z_a/(m_a Z_i)$ is sufficiently large,
light ions are accelerated much more efficiently than heavy ones.
The coated thin layer can be then accelerated forward.
It detaches from the foil, and moves as a whole
in the longitudinal direction.
The light ions within a small solid angle have a
quasi-monoenergetic energy spectrum.
The thinner the coating the narrower
energy spectrum of light ions is.

The important requirement is that
the transverse size of the coating
must be smaller than the laser waist
since an inhomogeneity in the laser pulse causes
nonuniform accelerating electric field and thus a
degradation of beam quality, as seen in experiments presented
in Refs. \cite{ions1} where the   exposed targets
had a thin proton layer on their surface.
The laser pulse inhomogeneity results in an additional energy spread of the ion
beam as seen in the experiments. The effect of the finite waist of the
laser pulse leads also to an undesirable defocusing of the fast
ion beam. In order to compensate for this effect and to focus the
ion beam, we can use properly deformed targets, as suggested in
Refs. \cite{Es99,W}.

In a target, whose transverse size is much greater than the laser
focal spot,
quasi-static electric field can be affected
by background `cold' electrons from the periphery.
To increase the quality and life-time of the quasi-static electric field,
we should use targets with transverse size less than the laser waist.

In order to estimate the typical energy gain of fast ions, we
assume that the main portion of the free electrons produced by ionization in
the irradiated region of the foil are expelled. In this case the
generated electric field near the positively charged layer is equal to
$E_{0} = 2\pi n_{0}e Z_{i} l$. Here $n_{0}$ is the ion density
and $e Z_{i}$ is an average ion electric charge in the foil,
and $l$ is the foil thickness.
The region of strong electric field has a transverse
size of the order of the diameter $2R_{\perp}$ of the focal spot.
Thus
the longitudinal size of the region where the electric field
remains essentially one-dimensional is of the order of
$R_{\perp }$
and the typical energy of an ion with charge $e Z_{a}$
accelerated by this electric field can be
estimated as ${\cal E}_{max} = 2\pi n_{0}Z_{a}Z_{i}e^{2}lR_{\perp}$.
We have assumed that the electron energy in the
laser field is well above the ion energy and larger than the
energy required for the electrons 
to leave the irradiated region. 
The maximum
electron energy in the electromagnetic wave is given by ${\cal
E}_{e} = m_{e}a^{2}/2$, where $a=eE/(m_{e}c\omega) $ is the
dimensionless amplitude of the laser pulse. From this condition we
can find the required value of the laser pulse intensity and its
power.

Consider a double layer target in the form of a
prolate ellipsoid coated at the rear side
with a very thin proton layer.
We assume that ellipsoid semiaxes are $R_\perp$ and $l/2$.
We can estimate the energy spectrum of protons,
using formulae for the electric field of the
electrically charged prolate ellipsoid \cite{LLEDCM}.
Let the $x$-axis be in the longitudinal direction, originating
from the target center.
On this axis the $x$-component of the electric field is given by
$E_{x}(x) = (E_{0}/3) R_{\perp}^{2}/(R_{\perp}^{2}-(l/2)^{2}+x^{2})$.
The
distribution function of the
fast protons $f(x,v,t)$ obeys the kinetic equation
$\partial_{t}f+v\partial_{x}f+\left( eE(x)/m_{p}\right) \partial _{v}f=0$.
If particles trajectories do not intersect or self-intersect
(for our ellipsoidal target this it true),
we have $f(x,v,t)=f_{0}(x_{0},v_{0})$,
where $f_{0}(x_{0},v_{0})$ is the distribution
function at the initial time $t=0$.
The
number of particles per unit volume in
phase space, $dxdv$, is equal to $ dn=fdxdv=fvdvdt=f\ d{\cal E\ }dt/m_{p}.$ We
assume that at $t=0$ all particles are at rest and that their spatial
distribution is given by $ n_{0}(x_{0})$ which corresponds to the distribution
function $ f_{0}(x_{0},v_{0})=n_{0}(x_{0})\delta (v_{0})$, with $\delta
(v_{0})$ the Dirac delta function. For the applications discussed in this paper
we are interested in the particle distribution function integrated over time.
Time integration of the distribution $fvdvdt$ gives the energy spectrum of the
beam $N({\cal E})d{\cal E}=\left( n_{0}(x_{0})/m_{p}\right) \left|
dt/dv\right|_{v=v_{0}}d{\cal E}.$ Here the Lagrange coordinate of the particle
$x_{0}$ and the Jacobian $\left| dt/dv\right| _{v=v_{0}}$ are functions of the
particle energy ${\cal E}$. The Lagrange coordinate dependence on the energy
$x_{0}=x_{0}({\cal E})$ is given implicitly by the integral of the particle
motion: ${\cal E}(x,x_{0})={\cal E}_{0}+e[\varphi (x)-\varphi (x_{0})]$, where
$\varphi (x)$ is the electrostatic potential. In the case under consideration, we
have ${\cal E}_{0}=0$ and $x=\infty $. The Jacobian $\left| dt/dv\right|
_{v=v_{0}}$ is equal to the inverse of the particle acceleration at $t=0$, i.e.
$\left| dt/dv\right| _{v=v_{0}}=1/|eE_{x}(x_{0})|$. On the other hand the 
function $|dt/dv|_{v=v_{0}}$ is equal to $|dx_{0}/d{\cal E}|$. Hence, we obtain
the expression for the energy spectrum in the form
\begin{equation} N({\cal E})d{\cal E}=
n_{0}(x_{0})d{\cal E}/|d{\cal E}/dx_{0}|_{x_{0}=x_{0}({\cal E})}.  \label{N2}
\end{equation}
We notice that the expression for the energy spectrum follows
from the general condition of particle flux continuity in the
phase space.

As we can see, in the vicinity of the target the
electric field on the axis is homogeneous and equals to
$E_{x}(l/2) = 2\pi n_{0}Z_{i}el/3$.
Therefore, the form of the energy spectrum
(\ref{N2}) is determined from the distribution of the proton density
$n_{0}(x=\varphi ^{-1}({\cal E}/e))$. We see that in general a
highly monoenergetic proton beam can be obtained when the function
$n_{0}(x_{0})$ is a strongly localized function, i.e. when the
thickness of the proton layer $\Delta x_{0}$ is sufficiently
small, and then $\Delta{\cal E}/{\cal E}\approx\Delta x_0/R_\perp$.

The longitudinal emittance of the beam is defined as the product
of the energy spread $\Delta {\cal E}$ and the time length $\Delta
t$ of the beam: $ \epsilon _{||}=\Delta {\cal E}\Delta t$.
Using the expressions obtained above we find the longitudinal emittance
of the accelerated proton beam
$\epsilon_{||}=\left( \Delta x_{0}/R_{\perp }\right) ^{2}
\sqrt{m_{p}{\cal E}R_{\perp }^{2}/2}$.
For ${\cal E}=100$MeV, $R_{\perp }=5\mu$m
and $\Delta x_{0}/R_{\perp }=6\cdot 10^{-3}$ we obtain
$\epsilon_{||}\approx 1.3\cdot 10^{-4}$MeV~ps which
is about 150 times
smaller than the emittance observed in the experiments with
non-optimized targets in Ref. \cite{Roth02}.

Near the axis, the radial component of the electric field  depends
linearly on the radius $r =\sqrt{y^{2}+z^{2}}$:
$E_r(r) = ( E_{0} l / 6R_{\perp}^{2}) r$.
We find that the particle
trajectory is described by $r = r_{0}\exp (\sqrt{kx}),$
where $r_{0}$ is the initial radial coordinate of the particle
and $k=l/R_{\perp }^{2}$.
From
this expression,    for
$l/R_{\perp }\ll 1$  we find the transverse emittance
$\epsilon_\perp=\pi r_{0}\Delta \theta $ of the fast proton beam:
$\epsilon_\perp=\pi r_{0}\sqrt{l/R_\perp}.$ Here
$r_{0}$ is the transverse size of the proton layer and $\Delta
\theta =\exp (\sqrt{kR_{\perp }})-1$ is the beam divergence angle.
For $r_{0}\approx 2.5\mu$m, $R_{\perp }=5\mu$m and
$l\approx 0.5\mu$m  the transverse emittance is of the order
of 2.5~mm~mrad.

For the hadron therapy
the particle flux must be approximately
$10^{10}$ to $5\cdot 10^{10}$ protons per second \cite{KhM}.
If the laser pulse is focused onto
a spot with diameter $10\mu$m,
CH layer with average density 1g/cm$^3$
and thickness $0.03 \mu$m
provides $10^{11}$ fast protons per pulse.

In order to take into account the numerous nonlinear and kinetic effects as
well as to extend our consideration to a multidimensional geometry, we
performed numerical simulations of the proton acceleration during the
interaction of a short, high power laser pulse with a two-layer target. We
used the three dimensional massively parallel and fully vectorized code REMP
(Relativistic Electro-Magnetic Particle-mesh code) \cite{REMP}. In these
simulations the largest number of grid cells was $2560\times 1024\times 1024$
and the number of quasiparticles was up to $820\cdot 10^{6}$. The boundary
conditions for the particles and for the fields are periodic
in the transverse directions and  correspond to  absorption
at the ends of the computation box along the $x$ axis.
The simulations were performed on 64 processors of the vector supercomputer NEC
SX-5 at CMC, Osaka University.

Here we present the results of these simulations.
The size of the simulation
box is $80\lambda \times 32\lambda \times 32\lambda $. A linearly polarized
laser pulse with  dimensionless amplitude $a=30$ propagates along the  $x$-axis.
The pulse size is  $15\lambda \times 12\lambda \times 12\lambda $. The pulse
has a trapezoidal shape (growth-plateau-decrease), with $3\lambda -2\lambda
-10\lambda $ in the   $x$-direction, and $1\lambda -10\lambda -1\lambda $
in the  $y$- and $z$-directions. The plasma consists of three species:
electrons, protons with $m_{p}/m_{e}=1836$, and heavy ions (gold with
$Z_{i}=+2$) with $m_{i}/m_{e}Z_{i}=195.4\times 1836/2$.

The first layer (gold) is placed at $x=5.5\lambda $. It has the
form of a disk with  diameter $10\lambda $ and  thickness
$0.5\lambda $. The second layer (protons) also has  the  form of a
disk with diameter $5\lambda $ and  thickness  $0.03\lambda $, and
is placed at the rear of the first layer, at $x=6\lambda $. The
electron density in the heavy ion layer corresponds to the ratio
$\omega _{pe}/\omega =3.0$ between the plasma and the laser
frequencies, for the proton layer it corresponds to
$\omega _{pe}/\omega=0.53$.
The number of electrons in the first layer is approximately
$2000$ times larger than that in the proton layer.

The simulation results are shown in Figs. \ref{fig:E} to \ref{fig:spectra},
where the
coordinates are measured in wave lengths of the laser light and
the time in laser periods. 
In Fig. \ref{fig:E} we present the electric field inside
the computation box, to show the shape of the transmitted laser pulse
and the longitudinal electric field accelerating the protons.
The electric field
is shown as a three-dimensional vector field;
it is localized in the vicinity of the first layer
(the layer of heavy ions) of the target and can be described as an electrostatic
field from a positively charged disk. The transmitted laser pulse is
presented by the isosurfaces of constant value of the
$z$-component of the electric field.
In Fig. \ref{fig:Z} we show the densities of plasma species
inside the computation box.
We see
that the proton layer moves along the $x$-axis and that the
distance between the two layers increases. The heavy ion layer
expands due to Coulomb explosion and tends to become rounded.
Part of the electrons is blown off by the laser pulse,
while the rest forms a
hot cloud around the target. We notice that for the
simulation parameters the electrons do not expelled from the region
irradiated by the laser light completely.
Even if only a portion
of the electrons is accelerated and heated by the laser pulse, the
induced quasi-static electric field appears to be strong enough to
accelerate the protons up $65~MeV$, as seen in
Fig. \ref{fig:spectra} presenting the energy
spectra of the protons and heavy ions.
The energy per nucleon
acquired by the heavy ions is approximately 380 times smaller than
the proton energy.
The heavy ions have a wide
energy spectrum while the protons form a quasi-mono-energetic
bunch with $\Delta {\cal E}/{\cal E}<5\%$.
We emphasize that the thinner the proton layer (or low-Z coating)
the better the beam quality is.
The proton beam
remains localized in space for a long time due to the bunching
effect of the linearly decreasing electric field in the acceleration direction.

In conclusion, the use of the multi-layer targets with different
shapes and compositions opens up new opportunities for controlling
and optimizing the parameters of the fast proton (ion) beam,
such as
the energy spectrum, the number of particles per bunch,
the beam focusing and the size of the region where the
beam deposits its energy.

\end{document}